\documentclass[prd,twocolumn,tightlines,showpacs,preprintnumbers,aps,floatfix,showpacs,amsmath,amssymb]{revtex4}
\usepackage{epsfig,amsmath,amssymb,bm,epsf,graphics}
\usepackage{dcolumn}

\newcommand{\be}{\begin{equation}}
\newcommand{\ee}{\end{equation}}
\newcommand{\ben}{\begin{eqnarray}}
\newcommand{\een}{\end{eqnarray}}

\begin{document}

\title{Gravitational lensing in the weak field limit by a braneworld black hole}
\author{A. S. Majumdar}
\email{archan@bose.res.in}
\author{Nupur Mukherjee}
\email{nupur@bose.res.in}
\affiliation{S. N. Bose National Centre for Basic Sciences, Block JD, 
Salt Lake, Kolkata 700 098, INDIA}

\date{\today}

\begin{abstract}
Braneworld black holes existing today may be of primordial origin, or
may even be produced in high energy particle
collisions in the laboratory and in  cosmic ray showers as well. 
These black holes obey a modified mass-radius 
relationship compared to standard Schwarzschild black holes.  Using the 
variational principle we calculate the bending angle of a 
light ray near the horizon of a braneworld black hole in the weak 
field limit. We next derive the expressions of several lensing quantities
like the Einstein radius and the magnification for a point light source.
These expressions are
modified compared to  the lensing quantities for standard Schwarzschild
black holes and contain the scale of the extra dimensions.
\end{abstract}


\maketitle

\section{Introduction}

There is widespread activity in braneworld gravity in
recent times\cite{maartens}. The braneworld scenario of our universe opens up
the fascinating possibility of the existence of large extra spatial
dimension(s)\cite{arkani}. The hugely popular Randall-Sundrum 
(RS-II) braneworld model\cite{randall} is made consistent
by the requirement that the
standard model fields are confined to the brane, except for gravity
which could also propagate into the Ads$5$ bulk which may be of infinite
extent but with curvature radius $l$. Current experiments 
probing the
resultant modification of the  Newtonian potential constrain the 
scale of the extra dimension to be $l \le 0.2 {\mathrm mm}$\cite{long}.

A specific issue of interest in braneworld gravity is the formation
and evolution of  black holes\cite{kanti}. Several types of black hole 
solutions have been obtained in the literature\cite{emparan,dadhich}. 
Black holes
formed due to horizon sized density perturbations in the modified
braneworld high energy phase of the early  universe
have a $5$-dimensional Schwarzschild metric.
The horizon size of such black holes is proportional
to the square root of their mass, a feature that modifies the Hawking
temperature, and consequently slows down the evaporation process\cite{guedens}.
It has been shown that these black holes accrete radiation in
the high energy phase which considerably prolongs their 
lifetimes\cite{majumdar}. 
Some of them could survive  up to the 
present era, thus acting as cold dark matter candidates. The braneworld
high energy phase is rather conducive to the formation of primordial
black hole binaries\cite{majumdar2}, and gravitational waves from such coalescing binaries are
likely to lie within the range of the next generation gravity
wave detectors\cite{inoue}.
On the other hand, super-horizon sized black holes which could be
formed by various collapse mechanism possess different 
geometries\cite{dadhich}, and might 
evaporate out rapidly as a consequence of Ads-CFT 
correspondence\cite{tanaka}. 

Since the $5$-dimensional fundamental scale
could be several orders of magnitude below the Planck scale, a lot of
current excitement stems from the possibility of braneworld black holes being
produced in high energy particle collisions\cite{giddings}.
Such a scenario would in principle, open up a direct experimental probe of 
extra dimensions and ($5$-d) Planck scale physics. This motivation has led to
several specific proposals for black hole formation in TeV scale dynamics in 
colliders such as the
LHC\cite{dimopoulos}. Black hole production in cosmic ray showers has
also been investigated with discussions on possible signatures\cite{feng}. 

The aim of the present paper is to investigate one potential avenue of
observational signatures for extra dimensions. The non-trivial 
spacetime curvature
around the vicinity of black holes could generate interesting motion for 
massless
and massive quanta passing near the horizon. Indeed, the bending
of light around standard black holes leads to the resultant observable
phenomenon of gravitational lensing which has been widely employed as a 
mechanism for detecting black holes in astrophysics\cite{eros}. The analysis of
light and particle motion in the
context of braneworld gravity is complicated by the absence till date of unique
analytical solutions for the metric representing compact objects in
higher dimensions. The modification of the standard $1/r$ form of the
Newtonian potential at small distances has nevertheless inspired several 
attempted solutions\cite{emparan} based on different physical requirements,
and certain corresponding analyses have been performed on the trajectory of
light rays and massive particles in such metrics\cite{wiltshire,frolov}. Some
interesting results on orbits around rotating $5$-dimensional black holes
have been derived\cite{frolov}. In this paper we explore
the lensing of optical sources by a braneworld black hole in the weak
field limit. We first
calculate the deflection angle for a light ray passing near the horizon
of a braneworld black hole. The corresponding lensing quantities are
derived next. Our analysis displays interesting departures of various
lensing phenomena and quantities compared to the standard Schwarzschild 
metric\cite{schneider}.

In our present work we consider a particular suggested geometry used
in Refs.\cite{guedens,majumdar,majumdar2,inoue} describing the 
spacetime metric near
the horizon of a $5$-dimensional braneworld black hole. 
Our analysis closely parallels
the derivation of the angle of bending of light in the Schwarzschild
metric using the variational principle for a null geodesic\cite{raychaud}.
Small
black holes formed with radius $r_h \le l$ in the early braneworld regime
could grow in size by accreting radiation\cite{majumdar,majumdar2}. Even
further growth to supermassive dimensions might be possible through
accretion of the background ``dark'' energy\cite{bean}. For such black holes
the metric far away from the horizon is expected to be of the standard
Schwarzschild form\cite{emparan2}. Our interest though is focussed on the 
region of
a few horizon lengths surrounding the black hole where departures from
the standard geometry could lead to interesting consequences, and
where the deflection of light is accounted for by the weak field
limit to a good degree of approximation. The metric in this region
is given by\cite{guedens}
\ben
dS_{4}^2= -\left(1-\frac{r_{h}^2}{r^2}\right)dt^2
+\left(1-\frac{r_{h}^2}{r^2}\right)^{-1}dr^2 \nonumber \\
+r^2\left(d\theta^2+\sin^2\theta d\phi^2\right),\label{1}
\een
where the horizon radius $r_h$ is related to the black hole mass $M$ by
\begin{equation}
 r_{h}^2 = \frac{8}{3 \pi} \left(\frac{l}{l_{4}}\right)
\left(\frac{M}{M_{4}}\right) l_{4}^2 \equiv {\cal P}M.
\label{2}
\end{equation}
with $l$ representing the size of the extra dimension, and $l_4$ and $M_4$
denoting the $4$-dimensional Planck length and mass, respectively. Note that 
Eq.(\ref{2})  signifies the altered mass-radius relationship for a braneworld
black hole (compared to $r^{\mathrm Sch}_h \sim M/M_4$ for the standard 
Schwarzschild
metric) and is at the root of the all the new results for light deviation
that follow from the braneworld metric. 

We are interested in the equation of motion for a light ray which is
described by a null geodesic   $d\tau^{2}= g_{ij}dx^{i}dx^{j}=0$. Thus,
in terms of an independent affine parameter $\lambda$, for a light ray
one could use Eq.(\ref{1}) and Eq.(\ref{2}) to obtain
\ben
-\left(1-\frac{{\cal P}M}{r^2}\right) \dot{t}^2 
+\left(1-\frac{{\cal P}M}{r^2}\right)^{-1}\dot{r}^2 \nonumber \\
+ r^2\left(\dot{\theta}^2+\sin^2\theta \dot{\phi}^2
\right)=0. \label{3}
\een
where we have defined
$\frac{dt}{d\lambda}=\dot{t}$, 
         $\frac{dr}{d\lambda}=\dot{r}$, 
         $\frac{d\theta}{d\lambda}=\dot{\theta}$ and
         $\frac{d\phi}{d\lambda}=\dot{\phi}$.  
Now, using variational principle one obtains the following equations:
\begin{equation}
\frac{d}{d\lambda}[\frac{dt}{d\lambda}(1-\frac{{\cal P}M}{r^2})]=0,\label{4}
\end{equation}
\begin{equation}
\frac{d}{d\lambda}[r^2 \sin^2\theta \frac{d\phi}{d\lambda}]=0,\label{5}
\end{equation}
\begin{equation}
\frac{d}{d\lambda}[r^2 \frac{d\theta}{d\lambda}]
=r^2\sin\theta\cos\theta (\frac{d\phi}{d\lambda})^2,\label{6}
\end{equation}
\ben
\frac{d}{d\lambda}[\dot{r} (1-\frac{{\cal P}M}{r^2})^{-1}]= -\frac{{\cal P}M 
\dot{t}^2}{r^3}
\nonumber \\
-(1-\frac{{\cal P}M}{r^2})^{-2} \frac{{\cal P}M}{r^3}\dot{r}^2 +r\dot{\theta}^2+r\sin^2\theta \dot{\phi}^2.\label{7}
\een

At this stage one could consider without loss of generality the orbit of
 the light ray
to be confined to the equatorial plane, i.e. $\theta=\frac{\pi}{2}$.
Using Eqs.(\ref{5}) and (\ref{6}) one obtains 
\begin{equation}
r^2\dot{\phi}=h.\label{8} 
\end{equation}
Further, using Eq.(\ref{4}) one gets
\begin{equation}
(1-\frac{{\cal P}M}{r^2})\dot{t}= k,\label{9}
\end{equation}
where $h$ and $k$ are  constants associated with respectively, the angular 
momentum and the energy of photons\cite{raychaud}.
Substituting Eqs.(\ref{8}) and (\ref{9}) in Eq.(\ref{3}) one gets
\begin{equation}
\dot{r}^2= k^2 - (1-\frac{{\cal P}M}{r^2})\frac{h^2}{r^2}.\label{10}
\end{equation}
Now in terms of the variable $u \equiv 1/r$ Eq.(\ref{10}) becomes
\begin{equation}
\frac{d^2u}{d\phi^2}= -u + 2 {\cal P}M u^3.\label{11}
\end{equation}

[The corresponding equation for the Schwarzschild metric is given by
$\frac{d^2u}{d\phi^2}= -u+3Mu^2/M_4^2$\cite{raychaud}]. 
Note that from Eq.(\ref{7}) one
could get an equation in terms of the variable $u$ given by 
\begin{equation}
\frac{d^2u}{d\phi^2}+ {\cal P}M u(\frac{du}{d\phi})^2+ u(1-\frac{{\cal P}M k^2}{h^2})=0.
\label{12}
\end{equation}
However, on combining Eq.(\ref{12}) with Eq.(\ref{11}) one is led to the
equation $\frac{d^2u}{d\phi^2} + 2u =0$ with no mass term
appearing in it. Hence, similar to the case
for the standard Schwarzschild metric\cite{raychaud}, the 
$r$-variation equation (\ref{7})
does not lead to any new input
on the calculation of the bending angle.
Without the mass term the solution of Eq.(\ref{11}) is given  by
$u=u_0{\mathrm cos}(\phi)$, with $u_0$ representing the inverse of the
distance of closest approach. Since the asymptote of the light path is
given by $u\equiv 1/r =0$, the solution $\phi=\pm\frac{\pi}{2}$, signifies no 
bending of light as is expected to be the case for vanishing mass.

In the weak field limit ($r \ge r_h$) the r.h.s of Eq.(\ref{11})  
introduces a small deflection of the equation of the light ray from
a straight line. The solution of Eq.(\ref{11}) with a small
contribution from the mass term can be written as 
\begin{equation}
u=u_{0}\cos(\phi)
+ {\cal P}M u_{0}^3[\frac{3}{4}\phi\sin(\phi)
-\frac{1}{16}\cos(3\phi)].\label{13}
\end{equation}
For small deflection $\delta$, one can substitute $\phi = \frac{\pi}{2}+\delta$
in Eq.(\ref{13}) and use the asymptote for the light ray ($u=0$) 
to obtain the total bending angle $\alpha=2\delta$ in the weak field
limit given by
\begin{equation}
\alpha = \frac{3\pi}{4} {\cal P}M u_{0}^2.\label{14}
\end{equation}
Note that the bending angle for the Schwarzschild metric which could be derived
in a similar fashion\cite{raychaud}  is given by
$\alpha^{\mathrm Sch} = 4 Mu_{0}/M_4^2$. The ratio of the bending
angles which is given by $\alpha/\alpha^{\mathrm Sch} = 2lu_0$, indicates
that the deflection by a braneworld black hole is more prominent for
small impact parameters $1/u_0 < l$. 

The expression for the 
bending angle of light $\alpha$ derived in the limit of a weak 
gravitational field follows
essentially due the the modified mass-radius relationship for a braneworld
black hole (Eq.(\ref{2})).
Note however, that 
the bending angle as a function of the impact parameter could also be obtained
from the general expression of deflection for a light ray in a spherically
symmetric metric\cite{weinberg}. This approach was used to calculate the
radius of the photon sphere $r_p^{\mathrm Sch} = 3M/M_4^2$ and the bending 
angle in the Schwarzschild metric\cite{virbhadra}. A similar analysis can
also be performed for braneworld black holes, and it turns out that the
the deflection angle can be expanded as
\begin{equation}
\alpha(u_0) = \frac{3\pi}{4} {\cal P} M u_{0}^2\biggl(1 + \frac{1}{2}{\cal P}M
u_0^2 + \frac{3}{8} ({\cal P}M)^2 u_0^4 + ...\biggr)
\label{15}
\end{equation}
for $1/u_0 > r_p$, with photon sphere defined by $r_p = (2PM)^{1/2}$. 
However, for small impact parameters the weak field approximation loses
validity. The discussion of strong gravitational field light bending 
for Schwarzschild
black holes incorporates special features such as the divergence of the
deflection angle and the winding number\cite{bozza}. We do not consider
such complications in our present analysis, and hence will restrict
ourselves to small bending angles given by Eq.(\ref{14}).

\begin{figure}[h!]
\begin{center}
\centerline{\epsfig{file=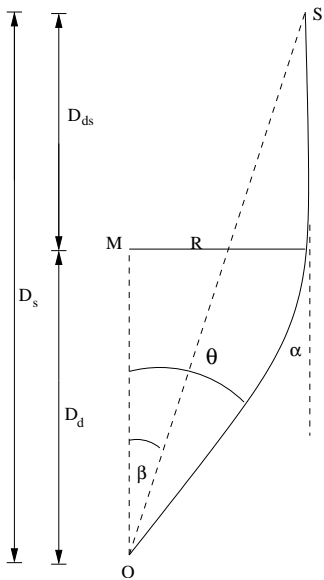,height=3.0in}}
\caption{Gravitationl lensing for point like mass object $M$. A light ray 
from the
source $S$ which passes the lens at a distance $R$ is deflected by $\alpha$. 
The observer sees an image of the source at angular position $\theta$.}
\end{center}
\end{figure}

The deflection of light causes the gravitational lensing of a light source
located behind the black hole for an observer. We now calculate various lensing
quantities like the Einstein angle and the 
magnification. For pointlike sources, we derive these expressions in an
approach that closely parallels the derivation of
similar lensing quantities for the standard Schwarzschild 
metric\cite{schneider}.
From Figure 1 it is evident that the condition that the light ray reach the 
observer is governed by the 
equation, $\beta D_{s}=\frac{D_{s}}{D_{d}} R - D_{ds} \alpha$.
Using the expression for the bending angle in Eq.(\ref{14}), one gets 
\begin{equation}
\beta =\theta - \frac{3\pi}{4}\frac{D_{ds}}{D_{s}}
\frac{{\cal P}M}{\theta^2 D_{d}^2}.\label{16}
\end{equation}
Defining
\begin{equation}   
\alpha_{0}^2 =\frac{3\pi}{4} 
\frac{D_{ds}{\cal P}M}{D_{s} D_{d}^2}\label{17}
\end{equation}
and
\begin{equation}
R_E = \alpha_{0}D_{d},\label{18}
\end{equation}
where $ \alpha_{0}$  and $R_{E}$ are the dimensionless  Einstein angle 
and the Einstein radius\cite{schneider},
respectively, Eq.(\ref{16}) becomes
\begin{equation}
\beta=\theta - \frac{\alpha_{0}^2}{\theta^2}.\label{19}
\end{equation}
This is the lense equation that is obtained for the braneworld 
metric (\ref{1}). 
It is instructive to compare Eq.(\ref{19}) with 
the Schwarzschild lens equation which is given by 
$\beta=\theta - (\alpha^{\mathrm Sch}_0)^2/\theta$ in which the Einstein angle
for the Schwarzschild metric is given by 
${\alpha}^{\mathrm Sch}_0 = \biggl(\frac{4MD_{ds}}{M_4^2D_{d}D_{s}}\biggr)^{1/2}$. 
The size of the Einstein radius corresponding to the braneworld metric
is much smaller compared  to the 
Einstein radius for the same mass in the Schwarzschild metric, i.e.,
\begin{equation}
\frac{R_E}{{R}^{\mathrm Sch}_E} = \Biggl(\frac{l}{2D_d}\Biggr)^{1/2}.\label{20}
\end{equation}

The image position $\theta$ can be obtained by 
solving Eq.(\ref{19}). 
Note that whereas the 
Schwarzschild lense equation is quadratic in theta, and hence has two
solutions, the braneworld lense equation (\ref{19}) is cubic in theta, thus
having only one real solution for the image position. The real root of 
Eq.(\ref{19}) is given by
\begin{eqnarray}
\theta_{1}=\frac{\beta}{3}+{2^{1/3}\beta^2\over 3[2\beta^3+27\alpha_{0}^2+3\sqrt{3}\alpha_{0}
(\sqrt{4\beta^3+27\alpha_{0}^2})]^{1/3}}
\nonumber\\
+ {[2\beta^3+27\alpha_{0}^2+3\sqrt{3}\alpha_{0}(\sqrt{4\beta^3+27\alpha_{0}^2})]^{1/3}\over 3 (2^{1/3})}.\label{21}
\end{eqnarray}
A special case arrises if the lens and the observer are colinear i.e. $\beta=0$
Then (\ref{19}) becomes
\begin{eqnarray}
\theta^3-\alpha_{0}^2=0\label{22}
\end{eqnarray}
whose real solution representing the Einstein ring\cite{schneider}  is 
given by $\theta=\alpha_{0}^{2/3}$.

The magnification $\mu$ produced at the image position $\theta$ is obtained
in terms of the source position $\beta$ by the relation\cite{schneider} 
\begin{eqnarray}
\mu = \frac{\triangle{\theta}}{\triangle{\beta}}
\frac{\theta}{\beta} = \vert \frac{\theta^4}{\theta^4 - \alpha_0^4}\vert
\label{23}
\end{eqnarray}
From the second equality in the expression for magnification
$\mu$ in Eq.(\ref{23}), one sees that $\mu$ is non-negligible only if
$\alpha_0 \sim \theta$. Using Eqs.(2) and (17), it follows that 
the impact parameter
$R$ should be of the order
$R \sim [(D_{ds}/D_s) (l/l_4) (M/M_4)]^{1/2} l_4$ for perceptible
magnification. The braneworld effect on lensing (i.e., the choice of
the metric (1) revealing the $5$-dimensional character of gravity at small
scales) is feasible for $R \leq l$, implying that $(M/M_4) \leq (l/l_4)$
if $(D_{ds}/D_s) \sim O(1)$. Choosing the value $(l/l_4) \sim 10^{30}$, 
one obtains the condition $(M/M_4) \leq 10^{30}$, or $M \leq 10^{-8} 
M_{\mathrm sun}$. Black holes with  such masses
have been classified as sub-lunar compact objects, and standard microlensing
results\cite{alcock} leave open the possiblity of their existence
in certain mass ranges as significant fractions of halo dark matter.
In the braneworld scenario, primordial black holes with such masses could
survive up to the present times\cite{majumdar}. It has been further argued that
braneworld black holes in such mass ranges could presently exist in the
form of binaries\cite{majumdar2} and that gravitational waves emitted 
during their coalescing stages
could be observed in future detectors\cite{inoue}.

In terms of the qauntities
\ben
 \beta^{\prime}=\frac{\beta}{\alpha_{0}}
\nonumber\\
\theta^{\prime}=\frac{\theta}{\alpha_{0}}
\label{24}
\een
and using Eq.(\ref{23}) the magnification by a braneworld black hole of
mass $M$ can be written as
\begin{eqnarray}
\mu &=&\frac{1}{3}\nonumber\\
&+&{2^{1/3}\beta^{\prime}[4+ \frac{3\sqrt{3}}
{\sqrt{4(\beta^{\prime})^3 {\alpha_{0}}+27}}]\over{9[2(\beta^{\prime})^3+\frac{27}{\alpha_{0}}+
\frac{3\sqrt{3}}{\alpha_{0}}
(\sqrt{4(\beta^{\prime})^3 {\alpha_{0}}+27})]}^{1/3}}
\nonumber\\
&+&{2^{2/3}(\beta^{\prime})^2[3+ \frac{3\sqrt{3}}
{\sqrt{4(\beta^{\prime})^3 {\alpha_{0}}+27}}]\over{9[2(\beta^{\prime})^3+\frac{27}{\alpha_{0}}+
\frac{3\sqrt{3}}{\alpha_{0}}
(\sqrt{4(\beta^{\prime})^3 {\alpha_{0}}+27})]}^{2/3}}
\nonumber\\
&-&{2^{4/3}(\beta^{\prime})^4[1+ \frac{3\sqrt{3}}
{\sqrt{4(\beta^{\prime})^3 {\alpha_{0}}+27}}]\over{9[2(\beta^{\prime})^3+\frac{27}{\alpha_{0}}+
\frac{3\sqrt{3}}{\alpha_{0}}
(\sqrt{4(\beta^{\prime})^3 {\alpha_{0}}+27})]}^{4/3}}
\nonumber\\
&-& {2^{5/3}(\beta^{\prime})^5[1+ \frac{3\sqrt{3}}
{\sqrt{4(\beta^{\prime})^3 {\alpha_{0}}+27}}]\over{9[2(\beta^{\prime})^3+\frac{27}{\alpha_{0}}+
\frac{3\sqrt{3}}{\alpha_{0}}
(\sqrt{4(\beta^{\prime})^3 {\alpha_{0}}+27})]}^{5/3}}
\nonumber\\
&+& {[2(\beta^{\prime})^3+\frac{27}{\alpha_{0}}+\frac{3\sqrt{3}}{\alpha_{0}}
(\sqrt{4(\beta^{\prime})^3 {\alpha_{0}}+27})]^{1/3}\over{9(2^{1/3})\beta^{\prime}}}.\label{25}
\end{eqnarray}
It can be checked that there is no magnification in the limit of zero mass,
i.e., $\mu = 1$ when $\beta^{\prime} \rightarrow \infty$. On
the other hand, in the limit
$\beta^{\prime} \rightarrow 0$, the magnification reduces to
\begin{eqnarray}
\lim_{\beta^{\prime}\rightarrow 0}\mu &=&\frac{1}{3} +
\frac{1}{3\beta^{\prime}\alpha_{0}^{1/3}}.\label{26}
\end{eqnarray}
Comparison with the magnification produced in the Schwarzschild metric shows
that in the 
(${\beta^{\prime} \rightarrow 0}$) limit,
\begin{equation}
\frac{\mu}{\mu^{\mathrm Sch}}\approx
\left(\frac{M_4}{M}\right)^{1/6}
\left(\frac{l}{l_{4}}\right)^{1/3}
\left(\frac{l_{4}}{D_{d}}\right)^{1/6}
\left(\frac{D_{s}}{D_{ds}}\right)^{1/6}\label{27}
\end{equation}
Note that a larger size of the extra dimension would produce a  brighter
image. The magnification produced by such braneworld black hole lenses which
could exist in our galactic halo, however, turns out to be diminished
compared to the standard Schwarzschild black holes, except for
extremely low masses. From Eq.(\ref{27}), and
using the value of $D = D_d D_ds/D_s \sim 10^{22} cm$ (relevant for
the lensing of galactic halo objects), and $(M/M_4) \sim 10^{30}$,
one obtains $(\mu/\mu^{\mathrm Sch}) \sim 10^{-4}$. In fact, one can see 
from Eq.(\ref{27}), that
in astronomical lensing Schwarzschild black holes
would produce brighter images than braneworld black holes for $M/M_4 > 10^6$.
Thus any braneworld black holes present in the galactic halo would be harder
to detect through lensing.

To summarize, in this paper we have investigated weak gravitational lensing 
of a point-like optical source by a
braneworld black hole. The study of particle and light motion in the
geometry of higher dimensional black holes is a subject of recent 
interest\cite{wiltshire,frolov}, particularly so because of the 
proposed mechanisms of black hole formation in high energy particle
collisions and cosmic ray showers\cite{giddings,dimopoulos,feng}.
Further, as has been shown recently\cite{majumdar}, braneworld black holes
could survive as relics from the early universe and act as candidates of
non-baryonic dark matter. It is feasible for 
primordial braneworld black
holes to exist in the form of binaries\cite{majumdar2}, and gravitational
waves from the coalescence of such binaries could be
detected in the near future\cite{inoue}. Thus, the exploration of the
phenomenon of gravitational lensing by braneworld black holes could be of
potential utility.

The geometry of a braneworld black 
hole incorporates a different mass-radius relationship\cite{guedens}
compared to a standard
Schwarzschild black hole. In the above analysis, we have
calculated the bending angle of light due to the gravitational 
potential of a braneworld black hole using the variational 
principle\cite{raychaud}.  The consisitency
of the derivation through the variational principle is confirmed
using the general expression for deflection angle in a spherically
symmetric metric\cite{weinberg}. The expression for the bending 
angle that we have derived contains the scale of the extra dimension $l$. 
We have next explored the phenomenon of gravitational lensing in the weak
field limit. The expressions of lensing quantities like the Einstein angle
and the magnification\cite{schneider} have been calculated in terms of 
the geometrical
parameters and the size of the extra dimension. The differences of these
expressions from the corresponding ones for Schwarzschild black hole
lensing\cite{virbhadra} have been highlighted. Further interesting
phenomena could be revealed through 
the analysis of strong gravitational lensing\cite{bozza} which in the context 
of the above
braneworld geometry has been recently worked out\cite{eiroa}. Strong
gravitational lensing in other braneworld metrics have also been 
studied\cite{whisker}.
Though our present observational capabilities might seem to restrict the
status of such analyses to theoretical curiosities, further improvement in 
techniques might enable the fascinating possiblity of discrimination of
different gravity models through observable 
lensing effects in the not too distant future.

\end{document}